\documentclass[journal, letterpaper, twocolumn]{IEEEtran}
\usepackage{multirow}
\usepackage[table]{xcolor}
\usepackage{adjustbox}
\usepackage{color}
\usepackage[T1]{fontenc}
\usepackage[latin9]{inputenc}
\usepackage{amsthm}
\usepackage{amsmath}
\usepackage{amssymb}
\usepackage{graphicx}
\usepackage{algorithmic}
\usepackage{algorithm}
\usepackage{array}
\usepackage{mathrsfs}

\theoremstyle{definition}

\theoremstyle{remark}

\usepackage{epstopdf}
\usepackage{setspace}
\usepackage{cite}

\begin{document}

\title{A Particle Filtering Approach for Enabling Distributed and Scalable Sharing of DSA Network Resources}

\author{Bassem~Khalfi,~\IEEEmembership{Student~Member,~IEEE,}
				Mahdi~Ben~Ghorbel,~\IEEEmembership{Member,~IEEE,}
				Bechir~Hamdaoui,~\IEEEmembership{Senior~Member,~IEEE,}
                Mohsen~Guizani,~\IEEEmembership{Fellow,~IEEE}
                and~Nizar~Zorba,~\IEEEmembership{Senior~Member,~IEEE}

\thanks{Bassem Khalfi and Bechir Hamdaoui are with Oregon State University, Oregon, USA (email: khalfib@eecs.orst.edu).}
\thanks{Mahdi Ben Ghorbel, Mohsen Guizani, and Nizar Zorba are with Qatar University, Doha, Qatar.}
\thanks{This work is an extended version of the paper published in IEEE Globecom Workshops (GC Wkshps) Dec. 2014~\cite{ghorbel2014resources}.}
}

\maketitle
\begin{abstract}
Handling the massive number of devices needed in numerous applications such as smart cities is a major challenge given the scarcity of spectrum resources. Dynamic spectrum access (DSA) is seen as a potential candidate to support the connectivity and spectrum access of these devices. We propose an efficient technique that relies on particle filtering to enable distributed resource allocation and sharing for large-scale dynamic spectrum access networks. More specifically, we take advantage of the high tracking capability of particle filtering to efficiently assign the available spectrum and power resources among cognitive users. Our proposed technique maximizes the per-user throughput while ensuring fairness among users, and it does so while accounting for the different users' quality of service requirements and the channel gains' variability. Through intensive simulations, we show that our proposed approach performs well by achieving high overall throughput while improving user's fairness under different objective functions. Furthermore, it achieves higher performance when compared to state-of-the art techniques.
\end{abstract}


\section{\sc {Introduction}}
\label{sec:introduction}
The increasingly growing number of wireless devices, along with the continually rising demand for wireless bandwidth, has created a serious shortage problem in the wireless spectrum supply. This foreseen spectrum shortage is shown to be due to the lack of efficient spectrum allocation and regulation methods rather than the scarcity of spectrum resources~\cite{FCC}.
As a result, DSA (Dynamic Spectrum Access) has been promoted as a potential candidate for addressing this shortage problem. This is by authorizing access to any unused spectrum opportunities by the non-legacy users~\cite{hamdaoui2009adaptive,Zhao2007}.
DSA embodies two main features: spectrum awareness and spectrum access. Spectrum awareness allows users to locate unused portions of the spectrum in all dimensions: time, frequency and space. For instance, one of the major challenging tasks encountered in DSA is to avoid harming Primary Users (PUs) with interference. Over the last decade or so, various different approaches have been proposed to identify spectrum opportunities (or holes) via spectrum sensing methods. Spectrum access is also of paramount importance since it encompasses the techniques for allocating and sharing these spectrum resources among the competing users, called Secondary Users (SUs)~\cite{el2012cooperative,Akyildiz2008}. By doing so, better spectrum efficiency is achieved, more users are served, and higher overall throughput is reached~\cite{ahmad2015survey}.

Despite the ever growing literature~\cite{Akyildiz2008,Tragos2013,ahmad2015survey}, the need for efficient spectrum allocation methods is still persisting especially with regards to the exponential growth of mobile devices and throughput demand. In fact, according to a recent study made by Cisco VNI mobile~\cite{Cisco}, mobile data traffic is anticipated to grow eightfold between $2015$ and $2020$, while the global number of mobile devices is expected to increase from almost $8$ to about $12$ billions. This anticipated growth raises new challenges, pertaining especially to scalability and resource allocation efficiency.

Broadly speaking, the problem of resource allocation could be addressed using either a centralized or a distributed approach. Although the former holds the promise to reach optimal solutions in some contexts as the spectrum manager or broker has a global view of the system, it suffers from serious scalability issues that essentially result from the huge amount of signaling overhead that can be involved. On the other hand, distributed approaches promise faster decisions and rapid adaptability to network variations, all of that through local information exchange, which makes them more scalable. All this is achievable but at the expense of having sub-optimal solutions. Furthermore, multichannel selection, when compared to single-band selection, promises higher data rates and better spectrum utilization. Smart cities are a potential application for distributed resource allocation schemes where sensors, e.g. smart cameras, are deployed to monitor sensitive activities (i.e. parking, traffic, security) and their data offloaded to data sinks. Typically, the number of sensors are an order of magnitude higher than the available number of bands, therefore intelligently assigning the bands among sensors is needed.

In the literature, various performance optimality criteria have been considered, including interference minimization, energy efficiency, throughput maximization, spectrum efficiency, delay minimization, etc. In this paper, we are mostly concerned with optimizing network throughput and users' fairness in a distributed, scalable manner.

\subsection{\sc {Related Works}}
\label{sec:LiteratureReview}
Resource optimization with various criteria has been the focus of numerous research works~\cite{ahmad2015survey,Ahmadi2012,Jiao2016,Duo2007,NoroozOliaee2013,Lei2010,guan2016distributed,Ren2016,el2016matching,ghorbel2016distributed,mwanje2016cognitive,Bazerque2007,Songtao2011}. For example, the optimal allocation of sub-carriers and modulation in OFDM DSA systems when considering centralized approaches was tackled in~\cite{Ahmadi2012}. The corresponding problem is NP-hard, thus the centralized approaches though optimal, are not practical due to their computational complexity limitation. To overcome this issue, two evolutionary algorithms, genetic algorithm and ant colony optimization, were proposed to give approximate/sub-optimal solutions. Using similar approach, the authors in~\cite{Jiao2016} relied on a two-tier crossover genetic algorithm. They targeted the maximization of the energy efficiency for DSA system with heterogeneous PU by deriving optimal allocation of power and bandwidth.

Since centralized approaches suffer from large signaling overheads, decentralized approaches with different objectives were considered instead~\cite{Duo2007,NoroozOliaee2013,ahmad2015survey,Lei2010,mwanje2016cognitive,guan2016distributed,Ren2016,ghorbel2016distributed,el2016matching}.
In~\cite{Duo2007}, the authors proposed an extended Kalman filtering-based adaptive game, where the DSA agents jointly decide on their transmission powers.
Similarly, the authors in~\cite{Lei2010} investigated the throughput maximization but for ad hoc cognitive networks by relying on the routing, dynamic spectrum allocation, scheduling, and transmit power control. In~\cite{ghorbel2016distributed}, the authors proposed a jointly distributed multiband spectrum and power allocation technique for large-scale DSA systems, based on reinforcement learning, that uses a new objective function that accounts for learnability, distributivity, and scalability. Authors in~\cite{mwanje2016cognitive} used reinforcement learning for self-organized cognitive cellular networks. The authors in~\cite{anandkumar2010opportunistic} considered cooperative spectrum allocation among secondary users using learning. However, this scheme suffers from a scalability issue as it requires the number of available channels to be higher than the number of users. Authors in~\cite{xu2012opportunistic} relaxed this constraint and proposed a learning technique, called SLA, to enable distributed spectrum assignment. While the proposed technique does not incur any signaling overhead between users, interference is not tolerated which restricts the access to only a few users.
In~\cite{Bazerque2007}, an online resource allocation targeting the maximization of the average sum-rate in an OFDMA-based cognitive network is proposed. This technique ensures fairness by imposing minimum rate requirements. This was achieved through the use of an ordinary sub-gradient method with slow convergence. The authors in~\cite{Songtao2011} improved the fairness by introducing to the link capacity expression the probability that a subcarrier is occupied.
It resulted in a faster convergence and higher capacity when compared to~\cite{Bazerque2007}. Authors in~\cite{cao2008distributed} introduced rules to regulate user behaviour and maximize the network fairness. However, physical layer aspects were not accounted for.
Scalability is the main shortcoming of these proposed approaches which was not studied especially when considering time varying channels.

Another possible method, that belongs to the family of stochastic search algorithms, is particle filtering (PF)~\cite{Doucet2000}. The core idea consists of estimating the conditional probability density through the use of the Monte Carlo simulations and the importance sampling techniques. This method has been proved to perform well in general scenarios without requiring extra-constraints on the model with comparison to Kalman-based filters (i.e., it works with non-linear models, non-Gaussian noise, multi-modal distributions, etc.). PF has shown its success especially with applications such as video tracking and localization~\cite{Gustafsson2010}. In wireless communications, it has been applied to blind equalization over frequency selective channels in SISO~\cite{Bordin2008} and multi-antenna systems~\cite{Du2005}. Moreover, it has been applied to signal detection~\cite{Yufei2004} and joint carrier recovery and channel estimation in OFDM systems over frequency selective fading channels~\cite{Hijazi2013}. However, very little effort has been put towards applying this technique to enable distributive DSA. For example, the authors in~\cite{Aggarwal2012} applied PF to devise a joint scheduling and power allocation technique for OFDMA-downlink systems. Specifically, they relied on PF to develop a greedy algorithm which aims to maximize an expected long-term goodput utility. The proposed algorithm achieves near optimal performance with practically reasonable computational complexity.

With all this in mind, there is still a need for developing efficient resource allocation methods for DSA systems that: i) are distributed so as to support large-scale DSA, ii) can achieve high per user throughput, and iii) ensure some level of fairness among users. These goals are, unfortunately, conflicting with one another. In our previous work~\cite{ghorbel2014resources}, we presented preliminary results on applying particle filtering theory for efficient distributed spectrum allocation. We evaluated the performance of our proposed technique in terms of total throughput and fairness in the single band case.

\subsection{\sc {Contributions}}\label{sec:Related-Work}
This paper proposes to solve a distributed joint multiband and power allocation in large-scale DSA systems using particle filtering theory. The distributivity of the proposed approach lies in the fact that channel selection decisions are made locally by each user without the need for a central entity. The proposed scheme strikes to achieve two global objectives, maximizing overall network throughput and ensuring per-user fairness, but in a distributed manner.
While deriving the optimal solution of the spectrum allocation problem is NP-hard~\cite{Ahmadi2012}, PF is shown to achieve a close-to-optimal solution in a distributed manner, thanks to its high tracking capability. Specifically, we show through simulations that when considering proportional fair objective function, particle filtering achieves a Jain's fairness index close to 1. In addition, stochastically modeling the problem allows to track the changes of the channel over time (the channels are time correlated).
To this end, our main contributions are summarized as follows:
\begin{itemize}
  \item Apply particle filtering theory to enable distributed resource allocation in large-scale DSA systems. The proposed distributed approach reduces the required information exchange and the processing delay at each node. We show that our technique achieves higher per-user average throughput when compared to reinforcement learning-based methods.

  \item Investigate the per-user throughput when considering multiband selection and compare it to the case of single-band selection subject to the same power budget. To the best of our knowledge, very limited works in the literature have addressed the multichannel multiradio allocation problem~\cite{Tragos2013}. The tradeoff between the number of selected bands and the resulted interference is also tackled in this paper.

  \item Consider and account for fairness metrics in our distributed resource allocation approach, and study the tradeoff between throughput and fairness performances.

\end{itemize}

Compared to our work~\cite{ghorbel2014resources}, this paper contains the following additions: $(i)$ supplementary analysis for the single band case, $(ii)$ study and analysis of the multiband assignment case, $(iii)$ study of the impact of primary user activity on the performance, and $(iv)$ study and analysis of the performance while considering  signaling overhead.
The remainder of this article is organized as follows. In Section~\ref{sec:System-Model}, we describe our system and channel model and discuss the different objective functions. In Section~\ref{sec:Opimal-Resource-Allocation}, we formulate the optimal resource allocation problem for large-scale DSA systems and discuss the issues related to the derivation of the optimal solution. We then present our proposed particle filtering based distributed DSA allocation technique in Section~\ref{sec:Single_Band}. Multiband spectrum allocation with primary user activity is introduced in Section~\ref{sec:Multi_band}. Simulation-based analysis and discussions are provided in Section~\ref{sec:simulation-results}. Finally, conclusions are drawn in Section~\ref{sec:conclusion}.

\begin{table}[!ht]
\caption{List of the Main Variables}
\begin{center}
\begin{tabular}{ |c|c| }
 \hline
 \textbf{Symbol} & \textbf{Notation} \\
 \hline
 $N$ & Number of users \\ \hline
    $m$ & Number of bands \\\hline
   $\textbf{v}(t)$ & Bands' availability vector \\\hline
    $R_n(t)$ & Throughput of user $n$ \\\hline
    $P_{n}(t)$ & Transmit power at user $n$ \\\hline
    $a$ & Channel allocation matrix \\\hline
    $\gamma_n$ & Received SINR \\\hline
    $N_0$ & Noise power density \\\hline
    $B^{(j)}$ & $j^{th}$ Channel bandwidth \\\hline
    $h_{ij}$ & Channel impulse \\\hline
    $\alpha_l$, $\xi$ & AR parameters\\\hline
    $r_{n}(t)$ & Reward at user $n$\\\hline
    $R_{n}^{th}(t)$ & Reward threshold at user $n$ \\\hline
    $\beta$ & Reward decaying factor \\\hline
    $\Theta_i(t)$ & Global objective function\\\hline
    $\mathcal{X}$ & State transition model \\\hline
    $\Psi$ & Observation model\\\hline
    $w_t^j$& Particle weights\\\hline
    $\omega$, $\textbf{u}(t)$ & AWG noise \\\hline
    $\ell$ &Maximum number of selected bands \\
 \hline
\end{tabular}
\end{center}
\end{table}

\section{\sc {Large-Scale DSA System Model}}
\label{sec:System-Model}
\begin{figure}
\begin{center}
\includegraphics[width=.9\columnwidth]{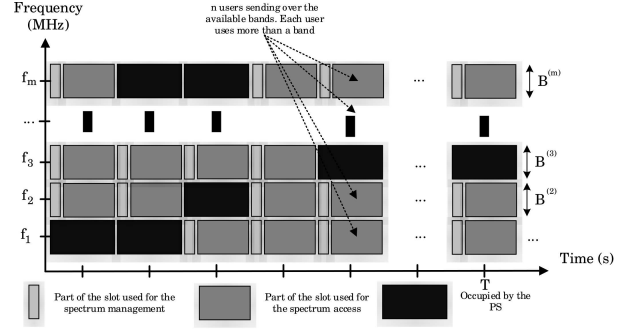}
\caption{The opportunities and occupancy of the spectrum in time and frequency. The spectrum is divided into $m$ non-overlapping bands. A part of each time slot is devoted for the spectrum sensing and for the allocation of the bands between the SUs.}
\label{sysmod}
\end{center}
\end{figure}

We consider a DSA system with $N$ transmitter-receiver pairs, all trying to communicate over a set of $m$ bands. We refer to a transmitter-receiver pair as a user throughout the paper. The $m$ bands have been sensed, where some bands are declared as occupied by the PUs and others are declared as available. An example of bands' occupancy is illustrated in Figure~\ref{sysmod}. To capture the PUs' activity, we introduce the availability vector $\textbf{v}(t)=[v_1(t),~v_2(t),... v_m(t)]$ where $v_k(t)=1$ if  band $k$ is available at time $t$ and $v_k(t)=0$ otherwise. This vector is updated at each time slot.
We assume that one multiband spectrum sensing technique is used at the beginning of each time slot to determine the availability of bands. This is motivated by the large number of proposed techniques in the literature to enable multiband spectrum sensing, e.g.~\cite{sun2013wideband,qin2016wideband} and references therein.

We consider the number of users to be very high compared to the number of the available channels ($N>> m$). We also assume that each user can communicate over multiple bands, and each user aims to achieve the maximum possible throughput $R_i(t)$, given its consumed power, $P_i(t)$, where $P_i(t)$ is expressed as
$P_i(t)=\sum_{j=1}^{m} a_i^{(j)}(t) P_i^{(j)}(t)$,
where $P_i^{(j)}(t)$ is the power allocated by user $i$ to the $j^{th}$ band and $a_i^{(j)}$ is a binary index indicating whether the $j^{\textrm{th}}$ band was selected by user $n$; here $\sum_{j=1}^{m}a_i^{(j)}\leq \ell$ where $\ell$ is the maximum number of bands selected by each user. 

The achieved throughput is expressed as
\begin{eqnarray}\label{eq:rate1}
R_i(t)=\displaystyle\sum_{j=1}^{m}v_j(t)a_i^{(j)}(t){B^{(j)}}\log_2(1 +\gamma_i^{(j)}(t)),
\end{eqnarray}
where $B^{(j)}$ is the $j^{th}$ channel bandwidth and $\gamma_i^{(j)}(t)$ is the received Signal-to-Interference plus Noise Ratio (SINR) of the $i^{th}$ user experienced at the $j^{th}$ band, which is expressed as
\begin{eqnarray}
\gamma_i^{(j)}(t)=\frac{P_i^{(j)}(t){|h_{ii}^{(j)}(t)|}^2}{\displaystyle{\sum_{\substack{k=1\\k\neq i}}^{N}}a_k^{(j)}(t)P_k^{(j)}(t){|h_{ik}^{(j)}(t)|}^2   +N_0B^{(j)}}
\end{eqnarray}
where $h_{ik}^{(j)}(t)$ is the $j^{th}$ channel impulse response from the $k^{th}$ transmitter to the $n^{th}$ receiver and $N_0$ is the power spectral density of the noise which is assumed to follow the gaussian distribution $\mathcal{CN}(0,N_0)$.

In order to capture the different users' requirements, instead of using as a local observation the achieved data rate $R_n(t)$, we propose to consider an elastic reward~\cite{ghorbel2016distributed}. Such reward function is more suited for web browsing and file transfer kind of data traffic and can be expressed as
\begin{equation}\label{elsrew}
r_i(t) = \begin{cases}
       R_i(t) & \text{if } R_i(t) >R_i^{\textrm{th}} \\
       R_i(t)\exp\big(-\beta\frac{R_i^{\textrm{th}}-R_i(t)}{R_i(t)}\big) & \text{otherwise},
       \end{cases}
\end{equation}
where $R_i^{\textrm{th}}$ is a rate threshold---i.e., user $i$ will not be satisfied had its received data rate been below $R_i^{\textrm{th}}$~\cite{Zorba2015}, and $\beta$ is a reward decaying factor.

We model the fading channel between a transmitter $k$ and a receiver $n$ by a $p^{th}$ order Auto-Regressive (AR($p$)) process~\cite{Duo2007}. Thus, at the time episode $t$, the channel impulse response $h_{ik}^{(j)}(t)$ is given by
\begin{equation}\label{eqnchannel}
h_{ik}^{(j)}(t)= \sum_{l=1}^{p}\alpha_l h_{ik}^{(j)}(t-l) + \xi \omega_i^{(j)}(t),
\end{equation}
where $\big\{\alpha_l\big\}_{l=1}^{p}$ and $\xi$ are the AR parameters that could be estimated using the Yule-Walker equations~\cite{Tsatsanis1996} which assumes that $\alpha_l=J_0(2\pi l f_d T_b)$, where $J_0$ is the zero$^{th}$ order Bessel function of the first kind, $f_d$ is the maximum Doppler frequency, $T_b$ is the channel coherence time, and $\omega_i^{(j)}(t)$ is an additive gaussian noise.

The main challenge addressed in this paper is how to assign the available bands among the $N$ users efficiently. Achieving this objective requires collaboration between the different users to gather information at a central unit which exploits this collected information to make centralized spectrum assignment decisions. Alternatively and in order to avoid the need for user collaboration, which often results in an excessive communication overhead, one can rely on users themselves to use local information to make these decisions in distributed manners. As mentioned earlier, examples of such distributed approaches are learning-based approaches in which users go after some defined objective functions to maximize their achieved throughput.
In the next section, we formulate our optimization problem and we discuss the need for a heuristic method to efficiently allocate the spectrum bands and the power among the different users.

\section{\sc {Optimal Resource Allocation in DSA}}
\label{sec:Opimal-Resource-Allocation}
We must first define the objective function targeted when allocating the different resources before formulating the problem. Thus, we start by discussing the different types of objective functions that we consider in this work.

\subsection{Global Objective Functions}
The authors in~\cite{NoroozOliaee2013} showed that the use of intrinsic objective functions results in fluctuating behaviors, whereas the use of global objective functions, which take into account other users' decisions, improve the overall system performance. These two objective functions (intrinsic and global) when corresponding to user $i$ can be expressed respectively as
\begin{equation}\label{intrinsic}
\Theta_i^{int}(t)=r_i(t)
\end{equation}
and
\begin{equation}\label{sum}
\Theta^{\textrm{sum}}(t)=\sum_{k=1}^Nr_k(t).
\end{equation}
A common problem with the above functions is that they do not ensure fairness among users. In an attempt to address fairness, a \emph{max-min} fairness approach, using a common global objective function known as bottleneck optimality, has been proposed in~\cite{Huaizhou2014} as
\begin{equation}\label{min}
\Theta^{\textrm{min}}(t)=\displaystyle{\min_{1\leq k\leq N}}\;\; r_k(t).
\end{equation}
This objective function is more suitable for users having the same requirements, which is generally not the case in wireless communications. Although this max-min approach solves the problem of starvation, it penalizes users with high requirements while giving users with low requirements more service than what they need. For a fairer allocation, proportional fairness~\cite{Dimitris2011} is shown to strike a good balance between the two conflicting objectives of maximizing total throughput and ensuring user fairness. The proportional fair function is defined as
\begin{equation}\label{pf}
\Theta^{\textrm{PF}}(t)=\sum_{k=1}^{N}\log (r_k(t)).
\end{equation}

In this work, we target the maximization of the average throughout, and thus we consider, out of the aforementioned objective functions, the global sum objective function given by Equation~\eqref{sum}. Besides, for users' fairness consideration we propose to consider the objective function expressed in Equation~\eqref{pf}.

Next, we formulate our resource allocation problem targeting the maximization of the two objectives: throughput and fairness.

\subsection{Optimal Resource Allocation}
Optimal allocation of spectrum and power resources among users can be achieved via centralized approaches, where users need to relay their information to a central unit which uses to solve a global optimization problem. Such a problem can be formulated as:
\begin{subequations}\label{eqn:optcent}
\begin{align}
\displaystyle\max_{}  &~\;\Theta_i(t)~~~~~~~~~~~~~~~~~~\;\forall\;i\in[1..N],~\forall\;t \label{subeq1}\\
\textrm{s.t.} &  ~\sum_{j=1}^{m} a_i^{(j)}(t)\leq\ell~~~~~~\,\forall\;i\in[1..N],~\forall\;t \label{subeq2}\\
&~ \sum_{j=1}^{m}P_i^{(j)}(t) \leq P_i^{\max}  ~~~\forall\;i\in[1..N],~\forall\;t \label{subeq3}\\
&~ P_i^{(j)}(t) \leq P_i^{(j),\max} ~~~~~~\;\forall\;i\in[1..N],\;\forall\;j\in[1..m], \label{subeq4}
\end{align}
\end{subequations}
where~\eqref{subeq1} is the global objective function to be maximized. In general, this problem is a mixed integer and real programming problem. The constraint~\eqref{subeq2} is used to control the number of the bands that the users could select at each scheduling time.
This is behind the combinatorial nature of the problem where each user is allowed to select up to $\ell$ bands.
To be in line with the current power requirements of wireless communication systems, we use the constraints~\eqref{subeq3} to limit the total consumed power, and~\eqref{subeq4} to limit the per-band consumed power.

Relying on a central unit, such as a spectrum broker or a fusion center which has a global view of the overall network, to solve this global optimization problem can be heavy computationally; this mixed integer and real programming problem is NP-hard. Mathematically speaking, the optimal solution could be found using an exhaustive search approach but in a non polynomial time~\cite{Aggarwal2012}. The computational complexity increases with the number of users, the number of bands, as well as the power levels.
The spectrum allocation only (without power allocation) has a computational complexity of approximately $\mathrm{O}(\eta^{N})$ with $\eta=(_{\ell}^m)$ and becomes infeasible when $N>>1$.
In addition, at every time slot $t$, each user should report the $N\times m$ channel gains, as well as its power budget and target throughput requirements, to its receiver and all the other receivers over all the $m$ bands. This can result in excessive control overhead.

To overcome these issues, heuristic approaches have been found more attractive, since they can find approximate (suboptimal) solutions in a reasonably acceptable computational time~\cite{Ahmadi2012}.
Among the benefits of using distributed approaches are memory savings and control message overhead reduction.

With all these issues in mind, we propose a distributed resource allocation algorithm based on particle filtering theory. Not only does our proposed approach achieve good approximate solutions in relatively acceptable time, but also reduces the delay resulting from the exchanging of channel reports. In what follows, we first start by single band spectrum allocation, and then multiband allocation.

\section{\sc {Single Band Distributed Spectrum Allocation}}
\label{sec:Single_Band}
In this section, we present our proposed particle filtering-based approach for single band spectrum allocation. Our optimization is formulated as follows
\begin{subequations}\label{eqn1}
\begin{align}
\displaystyle\max  &~~\;\Theta_i(t)~~~~~~~~~~\;\forall\;i\in[1..n],~\forall\;t \label{subeq11}\\
\textrm{s.t.} &  ~~\sum_{j=1}^{m} a_i^{(j)}(t)=\sum\textbf{a}_i(t)=1,~\forall\;i\in[1..n],~\forall\;t, \label{subeq21}
\end{align}
\end{subequations}
where $\Theta_i(t)$ is the global objective function described above, which depends on the channel selection $\textbf{a}_i(t)$ of each user. Although there is no power allocation, this is a non-linear integer programming problem. The constraint~\eqref{subeq21} is introduced to force each user to select at each time one single band. This is behind the combinatorial nature of the problem. Note that compared to Equation~\eqref{eq:rate1}, $\ell$ is set to be $1$ while in this part we do not consider the primary users' activity, i.e., $v_j(t)=1~\forall~j$.

One key merit that distributed resource allocation schemes possess is low signaling overhead. Local decisions are made following the exchange of some information (e.g. the achieved throughput and the selected band) among users and tracking the system evolution over time. In this context, particle filtering-based approaches are known to have strong tracking capabilities and can be adapted to non-linear and non-Gaussian estimation problems~\cite{Ondrej2013}. Hence, since the problem of spectrum assignment comes down to an estimation problem, we propose distributed particle filtering to estimate at each time slot the best spectrum allocation that achieves the set objective: joint network throughput and user fairness maximization.

\subsection{Particle Filtering for Spectrum Allocation}
The concept of distributed particle filtering is derived from the sequential estimation and importance sampling techniques. Each user needs to interact with some or all other users in order to get the best estimation of the unknown quantity, which represents in our system the spectrum allocation matrix $\textbf{a}(t)$. We model the evolution of the estimation of the best spectrum allocation as a discrete-time state-space model given by
\begin{subequations}\label{eqn:probmodel}
\begin{align}
 \textbf{a}(t)&= \mathcal{X}(\textbf{a}(t-1) )+ \textbf{u}(t),\label{subeq:state1}\\
 r_i(t)&=\Psi_i(\textbf{a}(t)) +z_i(t)\label{subeq:observation1},
\end{align}
\end{subequations}
where $\mathcal{X}(.)$ is a function that describes the state's change; $\Psi_n(.)$ is the function that links the global state $\textbf{a}(t)$ to the local observation $r_i(t)$, which is a non-linear function of the state $\textbf{a}(t)$; and $\textbf{u}$ and $z_i(t)$ are two stochastic noises of the state and the observation models, respectively. The noises are assumed to be white and independent of the past and the present states. Equation~\eqref{subeq:state1} describes the relation between the state at instants $t$ and $t-1$.
The local observation $r_i(t)$ represents the reward received by user $i$ when accessing the spectrum which is a function of the achieved throughput. The reward function maps the received throughput to a quality of experience as defined by Equation~\eqref{elsrew}.

The two equations~\eqref{subeq:state1} and~\eqref{subeq:observation1} provide a probabilistic model of our problem formulation. The goal of distributed particle filtering is to get the channel assignment matrix $\textbf{a}(t)$ sequentially using all the local measurements $r_i(t)$ of all users $n$ up until the current time $t$. From a practical point of view, each user should exchange information with only its neighbors or a relevant subset of users. In our work, we, however, assume that each user shares its measurements with all users similar to~\cite{NoroozOliaee2013}, as this is needed by the objective function $\Theta_i(t)$.

Since the channels' fading changes over time for the whole system, this affects the spectrum selection for each user at each time episode. Fortunately, with the presence of an inherent correlation between the channel realizations, the channel state at time $t$ could be estimated from the previous spectrum assignment; i.e., at time $t-1$. We assume that the users share their band selection, denoted as $\textbf{a}_n(t)$, along with its measured observation, $r_n(t)$, to the other users. This information allows them to estimate their best selections during the next time slot. Denoting the other users' band selections by $\textbf{a}_{-n}(t-1)$, the global function that governs the state change and executed by each user could be expressed as
\begin{eqnarray}\nonumber
\mathcal{X}(t)=
\arg {\displaystyle\max_{\textbf{a}_n(t)}} 
\Theta_i(t)|\{\textbf{a}_{-i}(t)=\textbf{a}_{-i}(t-1),\tilde{h}(t)\},
\end{eqnarray}
where $\tilde{h}$ is the estimate of the channel according to~\eqref{eqnchannel}.

With conventional Bayesian approaches, to estimate $\textbf{a}(t)$, we should compute the a posterior $f(\textbf{a}(t)|R_{1:N}(0:t))$, where $f$ denotes a probability density function and $r_{1:N}(0:t)$ is the vector that contains the observed throughput by all users from $t'=0$ until $t'=t$. The state can be sequentially estimated in two steps: a prediction phase given by Equation~\eqref{eq:pred} and an update phase given by Equation~\eqref{eq:up}~\cite{Ondrej2013} as
\begin{subequations}\label{eq:estim}
\begin{align}\nonumber
 &f(\textbf{a}(t)|r_{1:N}(0:t-1))= \\
 &~~~~~\int f(\textbf{a}(t)|\textbf{a}(t-1))f(\textbf{a}(t-1)|r_{1:N}(0:t-1)),\label{eq:pred}\\
  &f(\textbf{a}(t)|r_{1:N}(0:t))=\frac{ f(r_{1:N}(t)|\textbf{a}(t)) f(\textbf{a}(t)|r_{1:N}(0:t-1))}{ f(r_{1:N}(t))|r_{1:N}(0:t-1))}.\label{eq:up}
\end{align}
\end{subequations}
Although the recursion can simplify the derivation of the posterior $f(\textbf{a}(t)|r_{1:N}(0:t))$, it could not be straightforwardly computed due to the non-linearity and the involvement of an integral.

Particle filtering theory is an efficient tool to overcome this issue. Instead of calculating $f(\textbf{a}(t)|r_{1:N}(0:t))$, it suffices to consider a large number of samples from this distribution. These samples should be carefully drawn to reflect the original probability density function. Hence, it could be approximated by
\begin{equation}
f(\textbf{a}(t)|r_{1:N}(0:t))=\sum_{k=1}^{N_s}w^k(t)\delta(\textbf{a}(t)-\textbf{a}^k(t)),
\end{equation}
where $N_s$ is the number of samples, $\textbf{a}^k(t)$ is the $k^{th}$ sample and $w^k(t)$ is the correspondent weight. But, since we will apply the particle filtering distributively, instead of estimating $\textbf{a}(t)$, user $i$ estimates only its channel selection $\textbf{a}_i(t)$ by using a local density function known as importance density $f(\textbf{a}_i(t)|r_i(0:t),\textbf{a}_i(t-1),\textbf{a}_{-i}(t))$. {In this case, the particles, $\textbf{a}_i^k(t)$, are binary matrices composed by the other users' selections, $\textbf{a}_{-i}(t)$, and a possible band selection of user $i$ that corresponds to the $i^{th}$row.
User $i$ forwards its optimal selection $\textbf{a}_i(t)$ to the other users to be considered in their particles using a common control channel~\cite{kim2009distributed,aysal2009broadcast,cao2008distributed,noroozoliaee2013efficient}.
Note that this and other proposed approaches do allow the sharing of information among users, but not without incurring some unescapable overhead~\cite{kim2009distributed,aysal2009broadcast,cao2008distributed,noroozoliaee2013efficient}.

Although the importance density is optimal~\cite{Doucet2000}, its implementation is challenging, and hence, we instead utilize the following~\cite{Doucet2000,Ondrej2013}
\begin{equation}\label{impfun2}
\pi(\textbf{a}_i(t)|\textbf{a}(t-1))=f(\textbf{a}_i(t)|\textbf{a}^k(t-1),\textbf{a}_{-i}(t)).
\end{equation}
The weight at each sample is deduced from the previous weight and by taking into account the new observation. From the importance function in~\eqref{impfun2}, it follows that
\begin{equation}\label{weightfun2}
w_i^k(t) =w_i^k(t-1)~f(r_i(t)|\textbf{a}^{k}).
\end{equation}
These weights are then normalized.
\subsection{Single Band Spectrum Allocation: the Effect of the Number of Particles}

Over time, the weights of the different particles at each user become negligible, i.e., $w_i^k(t)\approx 0\;\forall\;k$ except for a few whose weights become very large. This problem is often known as the \textit{sample degeneracy}. And it implies that huge computations will be dedicated to update particles with very minor contributions. The idea of re-sampling is to make the particles with large weights more dominant while rejecting the ones with small weights~\cite{J}.

The number of particles is a key design parameter for the particle filtering algorithm as it affects both the computational complexity and the solution optimality. On the one hand, the larger the number of particles, the more accurate the approximation of the probability density function.
Furthermore, a large number of particles can result in a high computational complexity.
To assess this tradeoff, we study the effect of $N_s$ on the achievable throughput.
In Figure~\ref{Fig:numbparticle}, we plot the per-user average achieved throughput as a function of the number of particles using the intrinsic objective function, given by Equation~\eqref{intrinsic}.
Observe that increasing the number of particles beyond a certain value does not benefit the obtained throughput any further, yet results in increasing the computational cost. Clearly, the smaller the number of particles, the lower the computational cost, but if $N_s$ is chosen to be too small, it can lead to low throughput. The number that strikes a good balance between these conflicting objectives is between $10$ and $20$.
In the same figure, we also plot the per-user average throughput achieved when using the Q-learning approach~\cite{ghorbel2016distributed}, which does not, of course, depend on the number of particles. Observe that our proposed approach outperforms the Q-learning based approach.

\begin{figure}
\begin{center}
\includegraphics[width=1\columnwidth]{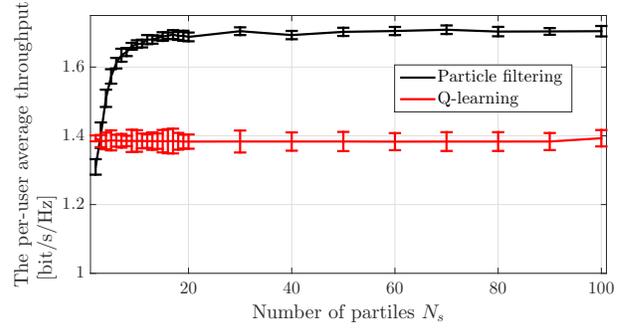}
\caption{The per-user average throughput when applying particle filtering with different number of particles using intrinsic objective function. The measured reward is after $T$ time episodes, with $T=20$. The number of users $n=200$ and the number of bands $15$.}
\label{Fig:numbparticle}
\end{center}
\end{figure}
A thorough evaluation of the performance of our proposed particle filtering-based approach vis-a-vis of its ability to maximize the per-user achievable amount of throughput and the level of fairness is provided later in Section~\ref{sec:simulation-results}.
Next, we consider spectrum and power allocation for the multiband scenario.

\section{Multiband Distributed Spectrum Allocation: Particle Filtering-Based DSA}
\label{sec:Multi_band}
Thanks to recent advances in wireless communication technologies, multiband access and sharing became possible with the emergence of cognitive radios. Although this promises higher throughput~\cite{Dong2011}, it could result in user's starvation and spectrum inefficiency if spectrum bands are not carefully assigned and shared among the different users. It is worth noting that existing distributed techniques are enabling multiband spectrum access. However, this comes with incurring some heavy cost often in terms of complexity and communication overhead~\cite{Wang2010}. We now present our distributed, yet simple DSA technique that relies on particle filtering theory to enable efficient distributed DSA without incurring much overhead.

Recalling our optimization problem in Equation~\eqref{eqn:optcent}, we propose to decouple the problem of spectrum allocation from that of power allocation. In a first phase, we use particle filtering to allocate spectrum among users, where each user is allowed to send over $\ell$ bands, and in the second phase, the power budget is distributed over the selected bands. For multiband spectrum selection, the same probabilistic model proposed by Equation~\eqref{eqn:probmodel} is used with the exception that the particle $\textbf{a}_i^k$ contains $\ell$ bands instead of one.

After selecting its bands, each user formulates the power allocation as:
\begin{subequations}\label{eq:powalloc}
\begin{align}
\max  &~~~~~~~~~\;\mathcal{R}_i(t)\\
&~~~~~~~~~ \sum_{j=1}^{m}a_i^j(t)P_i^{(j)}(t) \leq P_i^{\max}  \\
&~~~~~~~~~ P_i^{(j)}(t) \leq P_i^{j,\max} ~~~~~~~\;\forall\;j\in[1..m].
\end{align}
\end{subequations}
Note that the observation function $\Psi_i(\textbf{a}(t))$ could be written as
$\Psi_i(\textbf{a}(t))=\Phi(\mathcal{P}(\textbf{a}_i(t)),\textbf{a}_{-i}(t))$,
where $\mathcal{P}(.)$ denotes the weighted water filling algorithm applied for solving Equation~\eqref{eq:powalloc} and $\Phi(.,.)$ is the function relating the spectrum and power allocation to the observed reward.

The proposed algorithm is provided in Algorithm~\ref{algo:relgraph}.
\begin{algorithm}
\textbf{INPUT}: The power levels per user $\{P_i^{\max}\}_{1\leq i\leq N}$.\\
\textbf{OUTPUT}: The channel selection for every user $\{\textbf{a}_i\}_{1\leq n\leq N}$ and the power level at each channel\\
\begin{algorithmic}
\STATE \textbf{Initialization}
        At the first time slot $t_0$
        \FORALL{DSA user $i$}

        \FOR{$i=1:N$}
\STATE   \begin{enumerate}
               \item Generate random samples of the possible channel assignment $\textbf{a}_i^p(0)$;
               \item Set the weights to be equal $w_0^p=\frac{1}{N}$;
         \end{enumerate}
\ENDFOR
\ENDFOR

\FORALL{time slot $t$}
\STATE Perform the spectrum sensing and define the vector of the bands $\textbf{v}(t)$;
\FORALL{DSA user $n$}
\STATE   \begin{enumerate}			
               \item \textbf{Prediction:} Compute possible particles using~\eqref{impfun2};
			   \item \textbf{Decision:} Select $\ell$ bands of the particle giving highest reward;
               \item \textbf{Decision:} Allocate the power budget $P_n^{\max}$ among the selected $\ell$ bands;\\
			   \item Start the transmission on the selected bands;
    		   \item Update the channels estimation;
               \item \textbf{Weighting:}  Compute possible particles using~\eqref{weightfun2};
			   \item \textbf{Normalizing the weight:} $w_t^i=\frac{w_t^i}{\sum_{j=1}^N w_t^j}$;
	           \item \textbf{Re-sampling:} Apply re-sampling to avoid degeneracy.
         \end{enumerate}

\ENDFOR
\ENDFOR
\end{algorithmic}
\caption{Particle filtering based resource allocation for large-scale DSA system.}
\label{algo:relgraph}
\end{algorithm}

\section{Performance Evaluation}
\label{sec:simulation-results}
\subsection{System Setup}
We consider a DSA system with $N=200$ agents (an agent refers to a pair of nodes communicating with each other) communicating over $m=15$ bands unless specified otherwise. We assume that at the beginning of each time episode, the sensing process is performed and the available bands are determined.
The channels between the transmitter and its correspondent receiver as well as the other receivers are assumed to be Rayleigh fading channels with an average channel gain ${\Big[\frac{d_0}{d_{ki}}\Big]}^\eta$ where $d_0$ is a reference distance, $d_{ki}$ is the distance between the $i^{th}$ transmitter and the $k^{th}$ receiver and $\eta$ is the path-loss exponent set to $3$. We fix the average gain of the direct channel link to be $3~dB$ stronger than the average gains of the interference channels. To capture the channel correlation, the channels are used as a first order ($p=1$) AR process and the time coherence is chosen to be $T_b=1~ms$. The Doppler spread $f_d$ is caused by the mobility of the receiver at a maximum speed $v=70~Km/h$. Hence, the channel correlation over time $\alpha$ falls in the interval $[0.97,1]$. We assume that each user has a maximum transmit power $P_n(t)=3~dBm$ while the noise spectral density $N_0$ is set to $-100~dBm/Hz$.

We assume also that each user uses an elastic traffic model~\cite{ghorbel2016distributed}. In this model, a user $i$ has its own throughput requirement threshold, $R_i^{\textrm{th}}(t)$, which is uniformly distributed in the interval $[0,10kbit/s]$. If the value of some parameters is changed in the simulation, it will be stated so.

\subsection{System Performance}

We assess the effectiveness of the proposed resource allocation algorithm under two main scenarios: $(i)$ single band spectrum allocation without primary user activity and $(ii)$ multiband spectrum allocation with power allocation while considering primary user activity.

\subsubsection{Single Band Spectrum Allocation}

In Figure~\ref{throu_inter}, we investigate the per-user average throughput when considering intrinsic reward functions without primary user activity. In this scenario, given in Equation~\eqref{intrinsic}, particle filtering succeeds in tracking the channels' change and hence in selecting the best band that maximizes each user's reward. Also, when compared against the Q-learning-based allocation approach~\cite{ghorbel2016distributed} and SLA-based channel selection~\cite{xu2012opportunistic}, our proposed particle filtering approach achieves higher average throughput. Note that for SLA, we used the same parameters defined in~\cite{xu2012opportunistic} except $R_{m}$ is set to the per-user average throughput achieved by our scheme for fair comparison.
\begin{figure}
\begin{center}
\includegraphics[width=1\columnwidth]{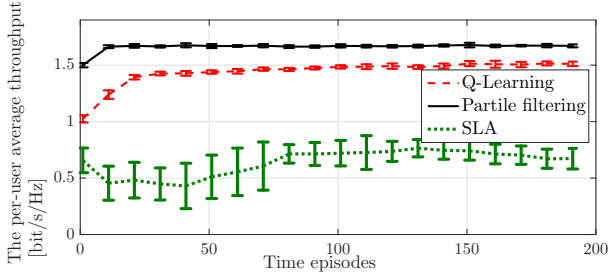}
\caption{Comparison between the per-user average throughput when using particle filtering based resource allocation with Q-learning~\cite{ghorbel2016distributed} and SLA~\cite{xu2012opportunistic}. The number of particles is $N_s=10$. The number of users $N=200$ and the number of bands is $m=15$. The used objective function is the intrinsic throughput reward.}
\label{throu_inter}
\end{center}
\end{figure}
In Figure~\ref{summin}, we compare the throughput performances obtained under each of the different objective functions, sum objective (Equation~\eqref{sum}), max-min fair objective (Equation~\eqref{min}), and proportional fair objective (Equation~\eqref{pf}), presented earlier in Section~\ref{sec:System-Model}. Observe that, as expected, the sum objective function achieves the highest average throughput among the three studied objective functions. Whereas, the proposed proportional fair function outperforms the max-min fair function. The reason behind the fact that the max-min fair function yields the least amount of throughput is that this function tends to penalize the users with good channels at the expense of favoring those users with poor channels to maintain the same level of throughput for all of them, thereby resulting in lesser total throughput, on the per-user average. As we will see next, this max-min fair function does, however, achieve good performances when it comes to fairness.

\begin{figure}
\begin{center}
\includegraphics[width=1\columnwidth]{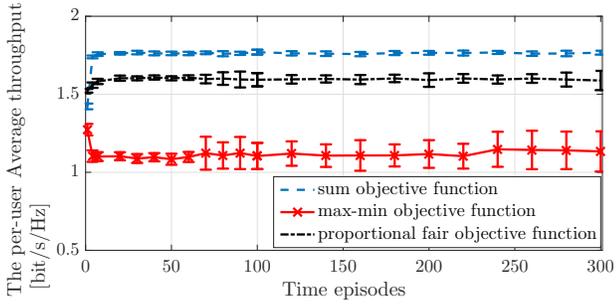}
\caption{Comparison of the per-user average throughput when applying particle filtering based spectrum allocation for different objective functions.}
\label{summin}
\end{center}
\end{figure}

We now assess how well our proposed scheme does vis-a-vis of its ability to ensure fairness among users, and we do so by measuring, plotting in Figure~\ref{Jain}, and comparing the Jain's fairness index~\cite{jain1984}, defined as
\begin{equation}\nonumber
J(t)=\frac{\big(\sum_{i=1}^Nr_i(t)\big)^2}{N\sum_{i=1}^Nr_i^2(t)},
\end{equation}
under each of the studied approaches.
First, observe that the proportional fair objective function achieves better fairness than the two other approaches, and the achieved fairness index is near optimal (very close to 1).
Second, the sum objective function has the lowest fairness index since its objective is to select the best channels that allow to reach the highest total throughput rather than accounting for users' satisfaction.
As shown by the performance behavior of the sum objective approach (when looking at both Figures~\ref{summin} and \ref{Jain}), it is clear that ensuring high average throughput
comes at the expense of not being fair to users, which is reflected in the low fairness index realized by the sum objective function. On the other hand, the max-min fair objective function, which is shown to obtain lower throughput than the sum objective function, achieves a better fairness index. Here, fairness is ensured at the expense of achieving lesser throughput. Unlike these two functions, our proposed proportional fair function allows to obtain the highest fairness index among the three studied techniques while achieving good throughput performances.

Using the same objective functions, we also plot the Jain's fairness index but when considering the Q-learning-based and SLA-based approaches  (see Figure~\ref{Jain} (bottom one)). Observe that when using the learning approach, the proportional fair objective function achieves better fairness but still not as good as when using the particle filtering-based approach. On the other hand, SLA achieves a high Jain's fairness index comparable to our technique when using the proportional fair objective function. However, recalling Fig.~\ref{throu_inter}, SLA does not achieve good per-user average throughput.
We therefore conclude that, when compared to the other techniques, our proposed particle filtering-based resource allocation technique coupled with our proposed proportional fair function, does strike a good balance between these two conflicting performance metrics: ensuring fairness among users and achieving high network throughput.

\begin{figure}
\centering{
\includegraphics[width=1\columnwidth]{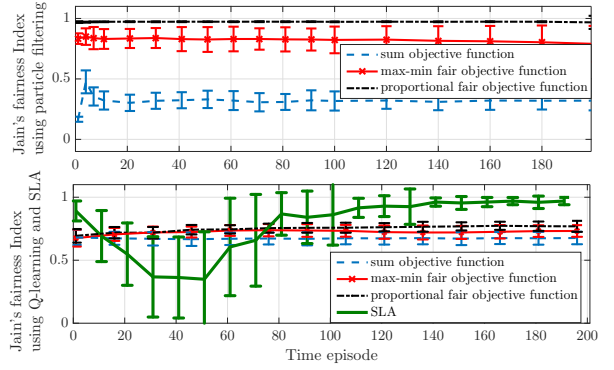}
\caption{Achievable Jain's fairness index under the studied schemes: sum throughput, min fairness and the proportional fairness using particle filtering, Q-learning-based resource allocation and SLA.}
\label{Jain}}
\end{figure}

\subsubsection{Multiband Spectrum Allocation with Primary User Activity}
\begin{figure}
\begin{center}
\includegraphics[width=1\columnwidth]{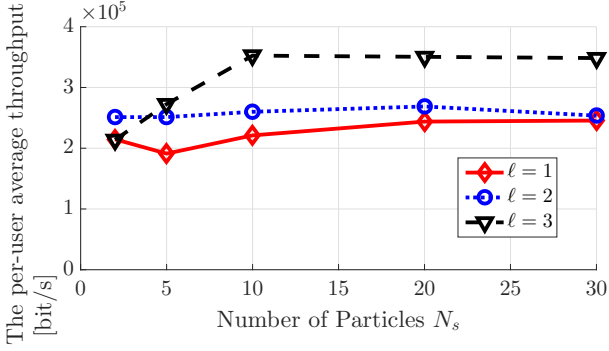}
\caption{The per-user average throughput when applying particle filtering for multiband selection with different number of particles using intrinsic objective function. The measured reward is after $T$ time episodes, with $T=10$. The number of users $N=100$, number of bands $m=10$ and bandwidth $B=1MHz$.}
\label{Fig:numbparticlemulti}
\end{center}
\end{figure}
In Figure~\ref{Fig:numbparticlemulti}, we study the effect of the number of particles when users are allowed to select multiple bands to access the spectrum. Two observations can be made from the figure. First, regardless of the number of bands, the per-user average achieved throughput performance increases with the number of particles until the number of particles reaches about $10$, after which the performance flattens out.
Second, when fixing the number of particles to $10$, the greater the number of bands, the higher the per-user achievable throughput.

Figure~\ref{multibandeffect} shows the effect of the number of selected bands by each user on the per-user average throughput. First, observe that for any given number of bands, the sum objective function outperforms the max-min and the proportional fair objectives due to the cooperative behavior between users, confirming thus the previous results. Second, as the number of selected bands increases, the achievable throughput first increases until it reaches a point where it starts to decrease, and this is regardless of the chosen objective function. The observed throughput performance degradation beyond a certain number of bands is mainly due to the effect of interference which is aggravated by the increase of the number of selected bands. Hence, this number should be optimized to reach a trade-off between achieving higher network throughput and minimizing user interference.

\begin{figure}
\includegraphics[width=1\columnwidth]{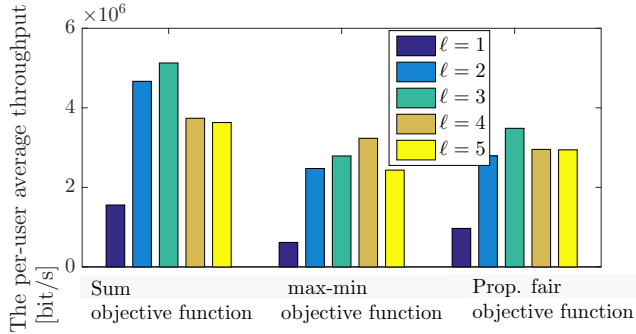}
\caption{The effect of multiband selection for $N=50$, $m=8$, $N_s=20$, and $B=6~MHz$.}
\label{multibandeffect}
\end{figure}

Having assessed the performance of our proposed technique in terms of throughput and fairness, we summarize these results in Table~\ref{tab:2} when compared to related works. In particular, we show that our proposed technique achieves higher performance with the tradeoff of higher signaling overhead.
\begin{table*}
  \centering
  \caption{Performance and signaling overhead comparison of the proposed scheme with related works.}\label{tab:2}
  \begin{adjustbox}{width=1\textwidth}
  \begin{tabular}{|l | c |c |c|}   
    \hline
    \textbf{Ref.}& \textbf{Performance}& \textbf{Shortcomings} & \textbf{System signaling overhead}\\  \hline  \hline
      \cite{anandkumar2010opportunistic}  & system throughput & $m>n$ (not scalable) &no signaling \\ \hline
      \cite{xu2012opportunistic} & system throughput and fairness & a maximum of $m$ users can access the spectrum at a time& no signaling \\ \hline
      \cite{cao2008distributed} & system fairness& no physical properties are conisdered & $O(n^2)$\\ \hline
      \cite{ghorbel2016distributed}&system throughput& low throughput and fairness& $O(n^2)$ \\ \hline
      \multirow {2}{*}{our work} & system throughput: sum& low fairness &$O(n^2)$ \\
                                                  & system fairness& high throughput and fairness&$O(n^2)$ \\ \hline
    \end{tabular}
    \end{adjustbox}
\end{table*}

As for primary users, we model their activity as an ON/OFF process, where each primary band is assumed to be occupied/busy with probability $p$. Assuming that all bands have the same bandwidth $B$, the occupancy probability is expressed as $p(t)=\sum_{k=1}^m(1-v_k(t))/m$.
In Figure~\ref{multibandprimaryactivity}, we consider multiband spectrum selection ($\ell=2$) for different primary user's activities. With the three objective functions, observe that the higher the primary user's activity is, i.e. the probability of occupancy $p(t)$, the lower the throughput each user achieves.

\begin{figure}
\includegraphics[width=1\columnwidth]{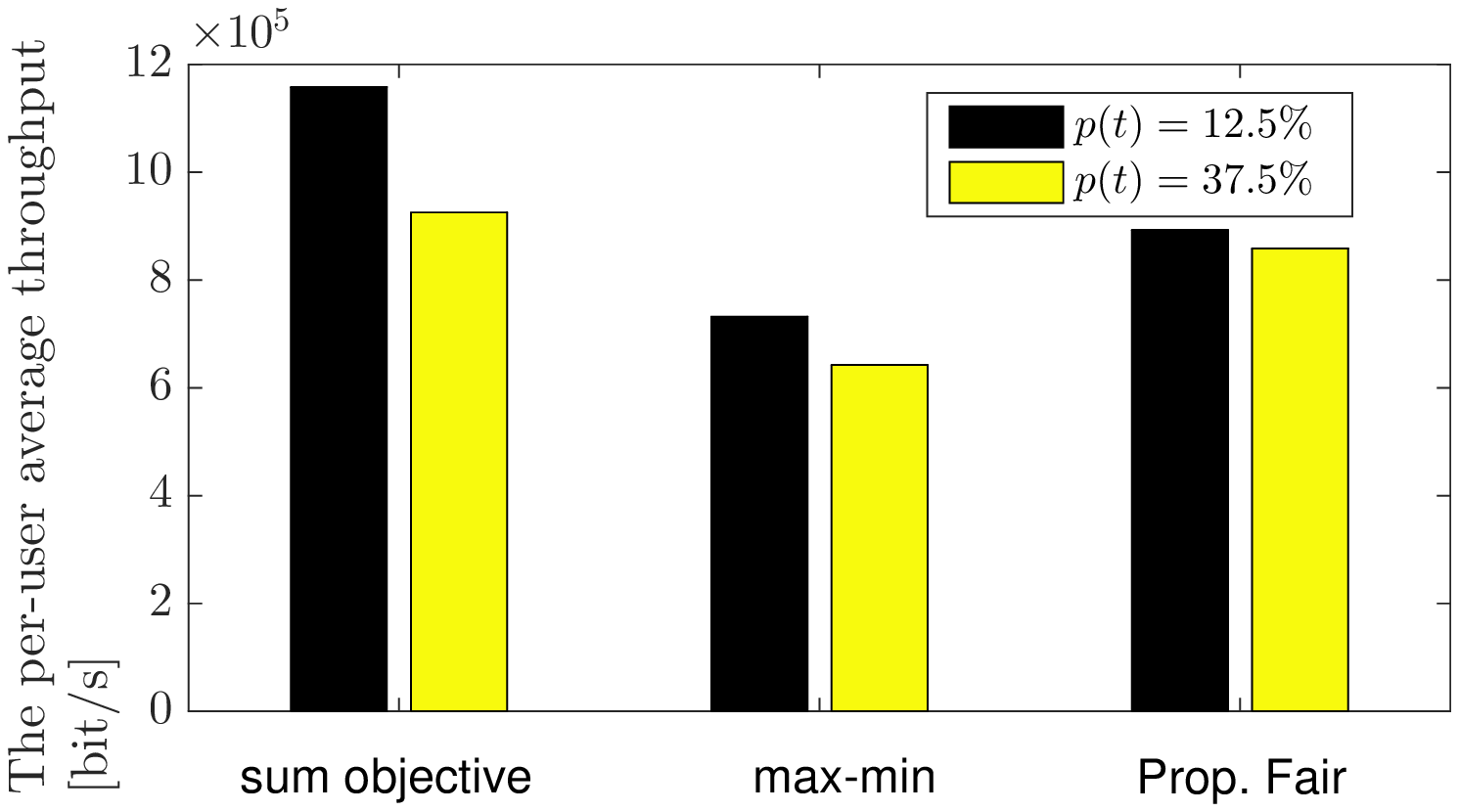}
{The primary user's activity effect for $N=100$, $m=8$, $N_s=20$, $B=6~MHz$, and $\ell=2$.}\label{multibandprimaryactivity}
\end{figure}

We investigate in Figure~\ref{powereffect} the impact of the power levels on the per-user average throughput. Simulations show that when each user is allowed to select a small number of bands ($\ell=2$), higher per-user average throughput is achieved as the transmit power increases. However, when the number of selected bands is high ($\ell=5$), the per-user average throughput drops regardless of the power level used. We conclude that sending over a large number of bands harms the system and cannot be fixed by controlling the transmit power.
 We also study the effect of the number of bands $m$ in Figure~\ref{bandeffect}. Observe that per-user average throughput increases when the number of considered bands increases for a given probability of occupancy $p(t)$ as less interference will be generated.  Also for a given number of bands $m$,  higher performance is achieved with the low PU activity as shown by Figure~\ref{multibandprimaryactivity}.
\begin{figure}
\includegraphics[width=1\columnwidth]{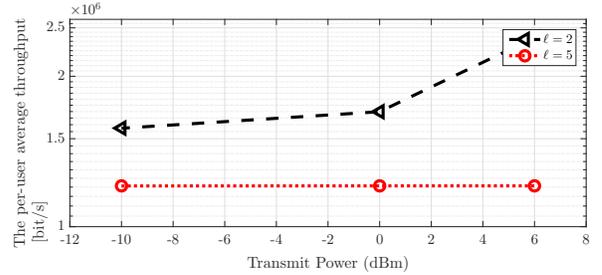}
\caption{The transmit power effect for $N=100$, $m=8$, $N_s=10$, $B=6~MHz$, and $p(t)=25\%$.}
\label{powereffect}
\end{figure}

\begin{figure}
\includegraphics[width=1\columnwidth]{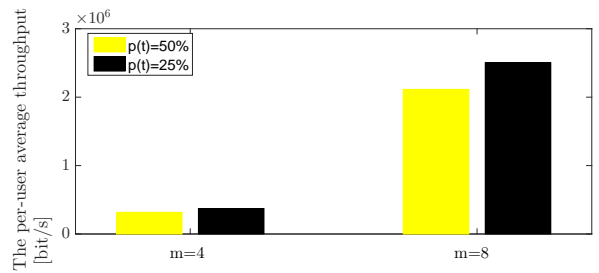}
\caption{The number of band effect for $N=100$, $N_s=10$, $B=6~MHz$, and $\ell=2$.}
\label{bandeffect}
\end{figure}

\section{Conclusions and Future Works}
\label{sec:conclusion}
This paper presents an efficient particle filtering based algorithm for a distributed multiband spectrum and power allocation in large-scale DSA systems. The performance of the proposed scheme was studied under different objective functions. The fairness and the per-user average throughput tradeoffs were studied. Furthermore, the effect of the number of the per-user selected bands and that of the primary users activities were investigated. The proposed approach is shown to achieve higher performance when compared to reinforcement learning based approaches with a relatively higher computational complexity.

\bibliography{References}
\bibliographystyle{ieeetr}
\end{document}